\documentclass[aps,prc,letterpaper,11pt,twoside,tightenlines,nofootinbib,showpacs,preprint]{revtex4}
\usepackage{graphicx}
\usepackage[sort&compress]{natbib}
\usepackage{subfigure}
\usepackage{amsmath}
\usepackage{amsfonts}
\usepackage{cancel}

\newcommand{\mev}{\textrm{ MeV}}
\newcommand{\be}{\begin{equation}}
\newcommand{\ee}{\end{equation}}
\newcommand{\ba}{\begin{eqnarray}}
\newcommand{\ea}{\end{eqnarray}}

\begin{document}

\title{A prediction of $D^*$-multi-$\rho$ states}

\author{C. W. Xiao $^1$, M. Bayar $^{1,\;2}$ and E. Oset $^1$}

\affiliation{$^1$Departamento de F\'{\i}sica Te\'orica and IFIC, Centro Mixto Universidad \\de Valencia-CSIC, Institutos de Investigaci\'on de Paterna, Apartado 22085, 46071 Valencia, Spain\\
$^2$ Department of Physics, Kocaeli University, 41380 Izmit, Turkey}

\date{\today}

\begin{abstract}
We present a study of the many-body interaction between a $D^*$ and multi-$\rho$. We use an extrapolation to SU(4) of the hidden gauge formalism, which produced dynamically the resonances $f_2(1270)$ in the $\rho\rho$ interaction and $D^*_2(2460)$ in the $\rho D^*$ interaction. Then let a third particle, $\rho$, $D^*$, or a resonance collide with them, evaluating the scattering amplitudes in terms of the Fixed Center Approximation of the Faddeev equations. We find several clear resonant structures above $2800\mev$ in the multibody scattering amplitudes. They would correspond to new charmed resonances, $D^*_3$, $D^*_4$, $D^*_5$ and $D^*_6$, which are not yet listed in the PDG, which would be analogous to the $\rho_3(1690)$, $f_4(2050)$, $\rho_5(2350)$, $f_6(2510)$ and $K^*_3(1780)$, $K^*_4(2045)$, $K^*_5(2380)$ described before as multi-$\rho$ and $K^*$-multi-$\rho$ states respectively.

\end{abstract}


\maketitle

\section{Introduction}

One of the important aims in the study of the strong interaction is to understand the nature and structure of hadronic resonances. The search for new resonances is another goal both in theories and experiments. At low energy, using the input of chiral Lagrangians \cite{Gasser:1984gg, Meissner:1993ah,Pich:1995bw,Ecker:1994gg,Bernard:1995dp} and implementing unitarity in coupled channels, one develops a theoretical tool, chiral unitarity theory, which explains the two-body interaction very successfully \cite{Kaiser:1995eg,Oller:1997ti,Oset:1997it,Oller:1998hw,Oller:1998zr, Oller:2000fj,Jido:2003cb,Guo:2006fu,Guo2006wp,GarciaRecio:2002td,GarciaRecio:2005hy,Hyodo:2002pk}. For the three-body interaction, the pioneer work of Ref. \cite{alberone} combined Faddeev equations and chiral dynamics and reported several S-wave $J^P=\frac{1}{2}^+$ resonances which qualify as two mesons-one baryon molecular states. One more step, using the Fixed Center Approximation \cite{Faddeev:1960su, Barrett:1999cw,Deloff:1999gc,Kamalov:2000iy} to Faddeev equations for multi-$\rho(770)$ states, was given in Ref. \cite{Roca2010} in which the resonances $f_2(1270)$, $\rho_3(1690)(3^{--})$, $f_4(2050)(4^{++})$, $\rho_5(2350)(5^{--})$, and $f_6(2510)(6^{++})$ were explained as basically molecules of an increasing number of $\rho(770)$ particles with parallel spins. Similarly, it was also found in Ref. \cite{Yamagata2010} that the resonances $K^*_2(1430)$, $K^*_3(1780)$, $K^*_4(2045)$, $K^*_5(2380)$ and a new $K^*_6$ could be understood as molecules made of an increasing number of $\rho(770)$ and one $K^*(892)$ meson. The $\Delta_{\frac{5}{2}^+}(2000)$ puzzle was explained in Ref. \cite{Xie:2011uw} with a $\pi-(\Delta\rho)$ system with the same method \cite{Xie:2010ig,Bayar:2011qj,Xiao:2011rc,Bayar:2012rk}. 

The Fixed Center Approximation to Faddeev equations is technically very simple and fairly accurate when dealing with bound states, as discussed in Ref. \cite{Bayar:2011qj}. It should also be avoided when dealing with states which have enough energy to excite its components in intermediate states \cite{MartinezTorres:2010ax}. It should be noted that the results of the Fixed Centre Approximation to Faddeev equations have been confirmed by the variational method approach with the effective one-channel Hamiltonian in a more recent work \cite{Bayar:2012dd} which predicts a narrow $DNN$ quasi-bound state in the energy range of about $3500\mev$.

In our present work, the main ideas of Refs. \cite{Roca2010, Yamagata2010} are followed to search $D^*$-multi-$\rho$ resonances in the charm sector. Using effective Lagrangians of the local hidden gauge theory \cite{hidden1,hidden2,hidden3,hidden4}, the $\rho\rho$ interaction was studied in Ref. \cite{Molina:2008jw} with on-shell factorized Bethe-Salpeter equations. It was found that the $\rho\rho$ interaction was attractive in the isospin zero, spin 0 and 2 channels, particularly in the tensor channel where it led to the formation of a $\rho\rho$ quasibound state or molecule that could be associated to the $f_2(1270)$ ( $I(J^{PC})=0(2^{++})$ ). With the same formalism, the composite systems of light ($\rho$ and $\omega$) and heavy ($D^*$) vector mesons were studied in Ref. \cite{Molina:2009eb}. In that work, a strong attraction was found in the isospin, spin channels $(I,S)= (1/2,0), (1/2,1)$ and $(1/2,2)$, with positive parity, leading to bound $\rho(\omega) D^*$ states, one of them identified as the $D^*_2(2460)\; (\;I(J^P)=\frac{1}{2}(1^-)\;)$. Therefore, the resonance $D^*_2(2460)$ was generated as a $\rho D^*$ quasibound state or molecule by the strong and attractive $\rho D^*$ interaction. As discussed in Ref. \cite{Roca2010}, because of the large binding energy per $\rho$ meson in spin 2, it is possible to obtain bound systems with several $\rho$ mesons as building blocks. As mentioned in Ref. \cite{Molina:2009eb}, the $\rho D^*$ interaction is also very strong and can bind the system. Thus the main aim in the present work is, first, to study the three-body interaction of $D^*$ and two $\rho$ mesons, for which we have two options, the clusters $D^*-f_2(1270)(\rho\rho)$ and $\rho-D^*_2(2460)(\rho D^*)$, in order to see if there are some resonance structures in the scattering amplitudes. If this is the case, one can predict a not-yet-discovered $D^*_3$ resonance, and we could continue our study extending these ideas to include more $\rho$ mesons as building blocks of the many-body system. Then we repeat the test in the four-body system and so forth. In analogy to the $K^*$-multi-$\rho$ systems \cite{Yamagata2010}, we find clear peaks in the scattering amplitudes for system with increasing number of $\rho$ mesons. Thus we predict new resonances in our work which are basically quasibound states or molecules made of an increasing number of $\rho(770)$ mesons and one $D^*$ meson which are not yet reported in the list of the PDG \cite{pdg2012}.

\section{$\rho\rho$ and $\rho D^*$ two-body interactions}

To evaluate the Faddeev equations under the Fixed Centre Approximation, we need to define the two-body cluster and then let the third particle collide with the cluster. Thus the starting point of our work is the two-body $\rho\rho$ and $\rho D^*$ interactions, which were studied in Refs. \cite{Molina:2008jw} and \cite{Molina:2009eb} with the local hidden gauge formalism \cite{hidden1, hidden2,hidden3,hidden4} and the unitary coupled channels method \cite{Kaiser:1995eg,Oller:1997ti, Oset:1997it,Oller:1998hw,Oller:1998zr,Oller:2000fj,Jido:2003cb,Guo:2006fu,GarciaRecio:2002td, GarciaRecio:2005hy,Hyodo:2002pk}. We briefly summarize the model of Refs. \cite{Molina:2008jw} and \cite{Molina:2009eb} here to explain how to obtain the unitarized $\rho\rho$ and $\rho D^*$ scattering amplitudes.

To evaluate the $\rho\rho$ and $\rho D^*$ scattering amplitudes with the coupled channels unitary approach, the Bethe-Salpeter equation in coupled channels is used:
\be 
t = [1-VG]^{-1} V,\label{BS}
\ee
where the kernel $V$ is a matrix of the interaction potentials in each channel, which is calculated from the hidden gauge Lagrangian. $G$ is a diagonal matrix of the loop function of every channel. The details can be seen in Refs. \cite{Molina:2008jw,Molina:2009eb}.

To construct the three-body system we start from the clusters $f_2(1270)$ ( $I(J^{PC})=0(2^{++})$ ) and $D^*_2(2460)\; (\;I(J^P)=\frac{1}{2}(1^-)\;)$ and add to them a $D^*$ or a $\rho$ respectively. The new particles are introduced with their spins aligned with that of the cluster such that the total spin adds one unity. Thus we only need to take into account the potential of spin $S=2$ for $\rho\rho$ and $\rho D^*$ interactions,
\ba 
V^{(I=0,S=2)}_{\rho\rho} (s) &=& -4 g^2 - 8 g^2 \Big( \frac{3s}{4m_\rho^2} - 1 \Big),\\
V^{(I=2,S=2)}_{\rho\rho} (s) &=& 2 g^2 + 4 g^2 \Big( \frac{3s}{4m_\rho^2} - 1 \Big),\\
V^{(I=1/2,S=2)}_{\rho D^*(11)} (s) &=& -\frac{5}{2} g^2 - 2\frac{g^2}{m_\rho^2} (k_1 + k_3) \cdot (k_2 + k_4) - \frac{1}{2}\frac{\kappa g^2}{m_\rho^2} (k_1 + k_4) \cdot (k_2 + k_3),\\
V^{(I=1/2,S=2)}_{\rho D^*(12)} (s) &=& \frac{\sqrt{3}}{2} g^2 + \frac{\sqrt{3}}{2}\frac{\kappa g^2}{m_\rho^2} (k_1 + k_4) \cdot (k_2 + k_3),\\
V^{(I=1/2,S=2)}_{\rho D^*(22)} (s) &=& \frac{1}{2} g^2 + \frac{1}{2}\frac{\kappa g^2}{m_\rho^2} (k_1 + k_4) \cdot (k_2 + k_3),\\
V^{(I=3/2,S=2)}_{\rho D^*} (s) &=& 2 g^2 + \frac{g^2}{m_\rho^2} (k_1 + k_3) \cdot (k_2 + k_4) + \frac{\kappa g^2}{m_\rho^2} (k_1 + k_4) \cdot (k_2 + k_3),
\ea
where $g=M_V/2f_\pi$ with $M_V$ the vector meson mass and $f_\pi$ the pion decay constant. In these equations $k_i,\; i = 1, 2, 3, 4$ are the initial, $(1,2)$, and final $(3,4)$ momenta of the particles. The quantity $\kappa = m_\rho^2/m^2_{D^*}$ appears because in some transitions one is exchanging a heavy vector instead of a light one. Note that in isospin $I=1/2$ there are two coupled channels, 1 is $\rho D^*$ and 2 is $\omega D^*$.

As mentioned in Refs. \cite{Molina:2008jw} and \cite{Molina:2009eb}, we also should take into account the contribution of the box diagram with two pseudoscalar mesons in the intermediate state. We only add the imaginary part of the box diagram contribution to the potential $V$, which is not accounted for by the coupled channels \cite{Wu:2010jy,Wu:2010vk}, and neglect the real part which is very small. We finally note that we take into account the $\rho$ mass distribution by replacing the $G$ function by its convoluted form by the mass distribution of the $\rho$ mesons in the loop. 

\begin{figure}
\centering
\includegraphics[scale=0.6]{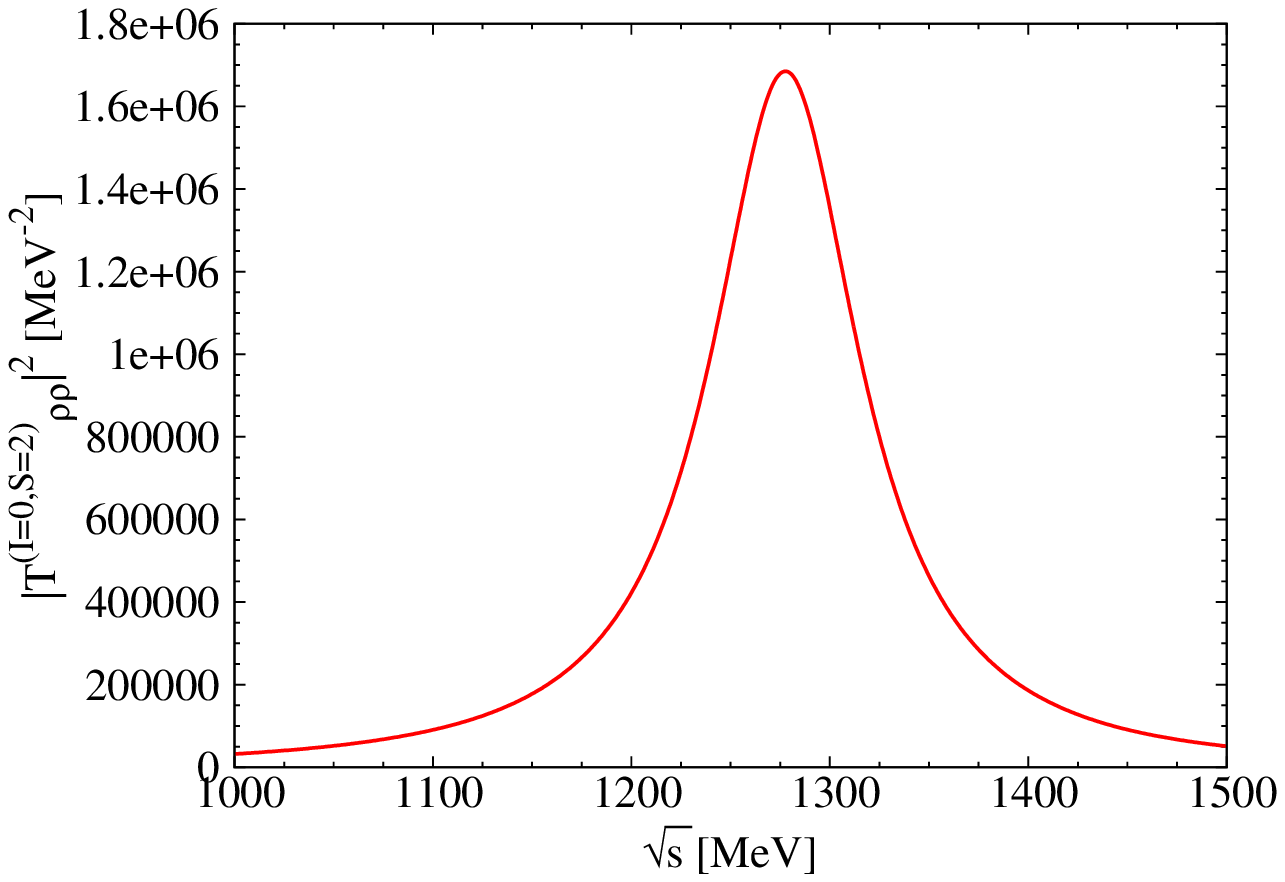}
\includegraphics[scale=0.6]{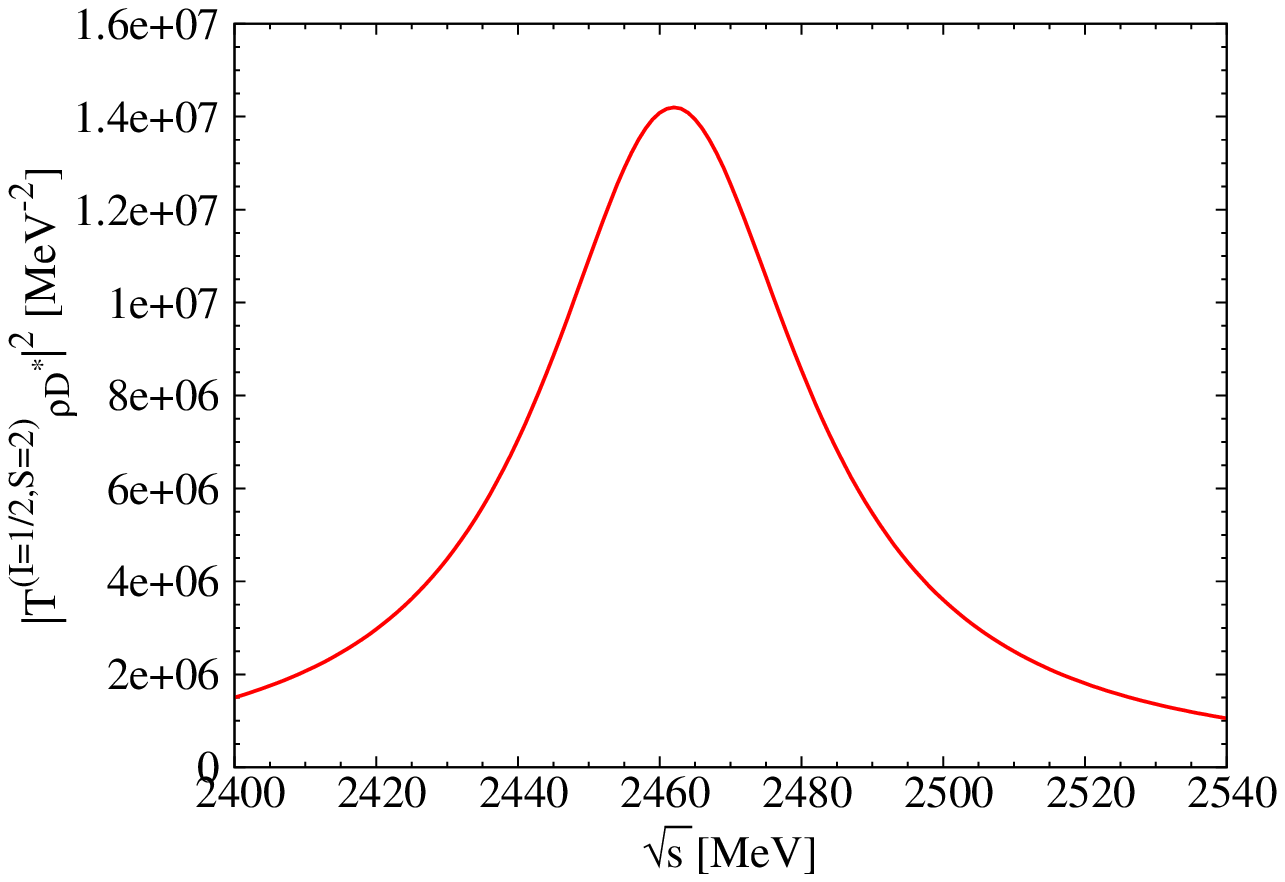}
\caption{Modulus squared of the scattering amplitudes. Left: $|t_{\rho\rho}^{I=0}|^2,\; f_2(1270)$; Right: $|t_{\rho D^*}^{I=1/2}|^2,\; D^*_2(2460)$.}\label{trhoD}
\end{figure}

Finally, considering the box diagram contribution to the potential $V$ and the convolution of the $\rho$ mass distribution in the loop function $G$, we show the evaluated results of $\rho\rho$ and $\rho D^*$ scattering amplitudes in Fig. \ref{trhoD}, which are consistent with Refs. \cite{Molina:2008jw} and \cite{Molina:2009eb}. The structure of the resonances $f_2(1270)$ and $D^*_2(2460)$ are clear in the peak of the modulus squared of the amplitudes. The nonresonant amplitudes $t_{\rho\rho}^{(I=0,S=2)}$ and $t_{\rho D^*}^{(I=3/2,S=2)}$ are not shown here.

\section{Multi-body interaction formalism}

\begin{figure}
\centering
\includegraphics[scale=0.4]{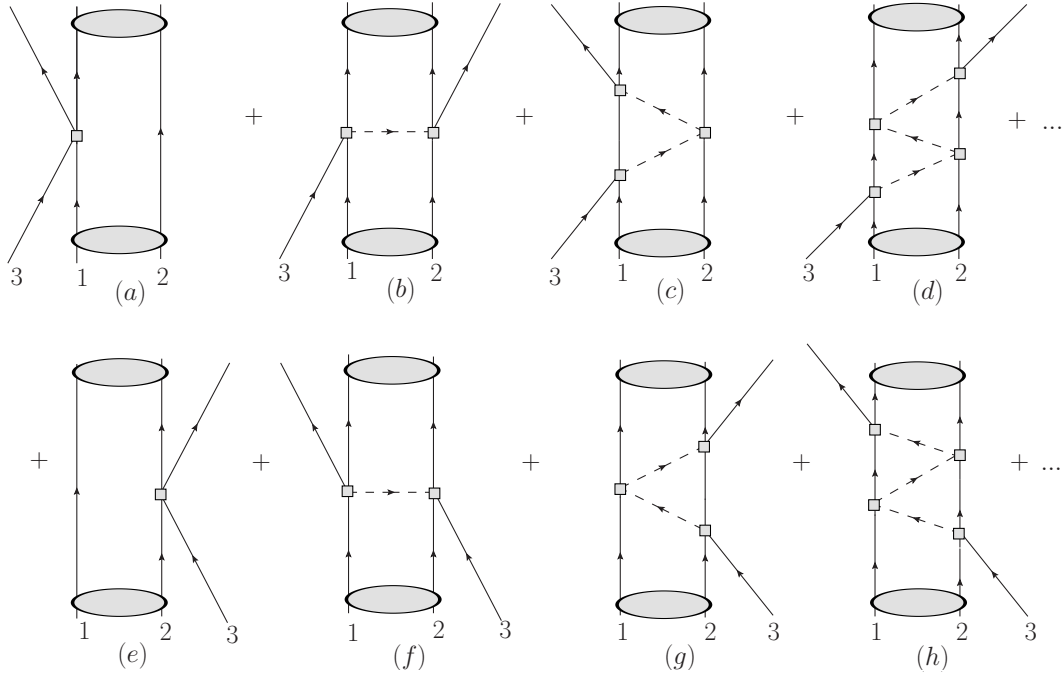}
\caption{Diagrammatic representation of the FCA to Faddeev equations.}\label{FCAfig}
\end{figure}

 The Faddeev equations under the Fixed Centre Approximation are an effective tool to deal with multi-hadron interaction \cite{Roca2010,Yamagata2010,Xie:2011uw,Xie:2010ig,Bayar:2011qj,Xiao:2011rc, Bayar:2012rk}. They are particularly suited to study systems in which a pair of particles cluster together and the cluster is not much modified by the third particle. The Fixed Centre Approximation to Faddeev equations assumes a pair of particles (1 and 2) forming a cluster. Then particle 3 interacts with the components of the cluster, undergoing all possible multiple scattering with those components. This is depicted in Fig. \ref{FCAfig}.
With this basic idea of the Fixed Centre Approximation, we could write the Faddeev equations easily. For this one defines two partition functions $T_1$, $T_2$ which sum all diagrams of the series of Fig. \ref{FCAfig} which begin with the interaction of particle 3 with particle 1 of the cluster ($T_1$), or with the particle 2 ($T_2$). The equations then read 
\begin{align}
T_1&=t_1+t_1G_0T_2,\label{threet1}\\
T_2&=t_2+t_2G_0T_1,\label{threet2}\\
T&=T_1+T_2,\label{threet}
\end{align}
where $T$ is the total three-body scattering amplitude that we are looking for. The amplitudes $t_1$ and $t_2$ represent the unitary scattering amplitudes with coupled channels for the interactions of particle 3 with particle 1 and 2, respectively. Besides, $G_0$ is the propagator of particle 3 between the components of the two-body system.
  
   For the unitary amplitudes corresponding to single-scattering contribution, one must take into account the isospin structure of the cluster and write the $t_1$ and $t_2$ amplitudes in terms of the isospin amplitudes of the (3,1) and (3,2) systems. Details will be discussed in next section. Besides, because of the normalization of Mandl and Shaw \cite{mandl} which has different weight factors for the particle fields, we must take into account how these factors appear in the single scattering and double scattering and in the total amplitude. This is easy and is done in detail in \cite{Yamagata2010,Xie:2010ig}. We show below the details for the present case of a meson cluster (also particle 1 and 2) and a meson as scattering particle (the third particle). In this case, following the field normalization of \cite{mandl} we find for the $S$ matrix of single scattering,
\begin{align}
\begin{split}
S^{(1)}_1=&-it_1 \frac{1}{{\cal V}^2} \frac{1}{\sqrt{2\omega_3}} \frac{1}{\sqrt{2\omega'_3}}
 \frac{1}{\sqrt{2\omega_1}} \frac{1}{\sqrt{2\omega'_1}}\\
 &\times(2\pi)^4\,\delta(k+k_R-k'-k'_R),\label{s11} \\
\end{split}\\
\begin{split}
S^{(1)}_2=&-it_2 \frac{1}{{\cal V}^2} \frac{1}{\sqrt{2\omega_3}} \frac{1}{\sqrt{2\omega'_3}}
 \frac{1}{\sqrt{2\omega_2}} \frac{1}{\sqrt{2\omega'_2}}\\
 &\times(2\pi)^4\,\delta(k+k_R-k'-k'_R),\label{s12} \\
\end{split}
\end{align}
where, $k,\,k'$ ($k_R,\,k'_R$) refer to the momentum of initial, final scattering particle ($R$ for the cluster), $\cal V$ is the volume of the box where the states are normalized to unity and the subscripts 1, 2 refer to scattering with particle 1 or 2 of the cluster.

  The double scattering diagram, Fig. \ref{FCAfig} (b), is given by
\be 
\begin{split}
S^{(2)}=&-i(2\pi)^4 \delta(k+k_R-k'-k'_R) \frac{1}{{\cal V}^2}
\frac{1}{\sqrt{2\omega_3}} \frac{1}{\sqrt{2\omega'_3}}
 \frac{1}{\sqrt{2\omega_1}} \frac{1}{\sqrt{2\omega'_1}}
 \frac{1}{\sqrt{2\omega_2}} \frac{1}{\sqrt{2\omega'_2}}\\
&\times\int \frac{d^3q}{(2\pi)^3} F_R(q) \frac{1}{{q^0}^2-\vec{q}\,^2-m_3^2+i\,\epsilon} t_{1} t_{2},\label{s2}
\end{split} 
\ee 
where $F_R(q)$ is the cluster form factor that we shall discuss below. Similarly the full $S$ matrix for scattering of particle 3 with the cluster will be given by
\be 
\begin{split}
S=&-i\, T \, (2\pi)^4 \delta(k+k_R-k'-k'_R)\frac{1}{{\cal V}^2}\\
&\times\frac{1}{\sqrt{2 \omega_3}} \frac{1}{\sqrt{2 \omega'_3}}
\frac{1}{\sqrt{2\omega_R}} \frac{1}{\sqrt{2\omega'_R}}.\label{sful}
\end{split} 
\ee

In view of the different normalization of these terms by comparing Eqs. \eqref{s11}, \eqref{s12}, \eqref{s2} and \eqref{sful}, we can introduce suitable factors in the elementary amplitudes,
\be
\tilde{t_1}=\frac{2m_R}{2m_1}~ t_1,~~~~\tilde{t_2}=\frac{2m_R}{2m_2}~ t_2,
\ee
where we have taken the approximations, suitable for bound states, $\frac{1}{\sqrt{2 \omega_i}}=\frac{1}{\sqrt{2m_i}}$, and sum all the diagrams by means of
\be 
T=T_1+T_2=\frac{\tilde{t_1}+\tilde{t_2}+2~\tilde{t_1}~\tilde{t_2}~G_0}{1-\tilde{t_1}~\tilde{t_2}~G_0^2}. \label{new}
\ee
The function $G_0$ in Eq. \eqref{new} is given by 
\be 
G_0(s)=\int \frac{d^3\vec{q}}{(2\pi)^3} F_R(q) \frac{1}{q^{02}-\vec{q}^{~2}-m_3^2 +i\,\epsilon }.\label{g0baryon}
\ee
where $F_R(q)$ is the form factor of the cluster of particles 1 and 2. We must use the form factor of the cluster consistently with the theory used to generate the cluster as a dynamically generated  resonance. This requires to extend into wave functions the formalism of the chiral unitary approach developed for scattering amplitudes. This work has been done in \cite{gamerjuan,yamajuan,Aceti:2012dd} for $s$-wave bound states, $s$-wave resonant states and states with arbitrary angular momentum respectively, here we are only need the expressions for $s$-wave bound states, and then the expression for the form factors is given in section 4 of \cite{yamajuan}, which we use in the present work and reproduce below 
\begin{align}
\begin{split}
F_R(q)&=\frac{1}{\mathcal{N}} \int_{|\vec{p}|<\Lambda', |\vec{p}-\vec{q}|<\Lambda'} d^3 \vec{p} \frac{1}{2 E_1(\vec{p})} \frac{1}{2 E_2(\vec{p})} \frac{1}{M_R-E_1(\vec{p})-E_2(\vec{p})} \\
&\quad\frac{1}{2 E_1(\vec{p}-\vec{q})} \frac{1}{2 E_2(\vec{p}-\vec{q})} \frac{1}{M_R-E_1(\vec{p}-\vec{q})-E_2(\vec{p}-\vec{q})}, \label{formfactor}
\end{split}\\
\mathcal{N}&=\int_{|\vec{p}|<\Lambda'} d^3 \vec{p} \Big( \frac{1}{2 E_1(\vec{p})} \frac{1}{2 E_2(\vec{p})} \frac{1}{M_R-E_1(\vec{p})-E_2(\vec{p})} \Big)^2, \label{formfactorN}
\end{align}
where $E_1$ and $E_2$ are the energies of the particles 1, 2 and $M_R$ the mass of the cluster. The parameter $\Lambda'$ is a cut off that regularizes the integral of Eqs. \eqref{formfactor} and \eqref{formfactorN}. This cut off is the same one needed in the regularization of the loop function of the two particle propagators in the study of the interaction of the two particles of the cluster \cite{yamajuan}. We take in the present work $\Lambda'=875\mev$, the same as used to generate the bound states \cite{Yamagata2010,Xie:2010ig} of $f_2(1270)$ in the two-body interaction, and $\Lambda'=1200\mev$ for getting the $D^*_2(2460)$ cluster. Thus we do not introduce any free parameters in the present procedure.
 
In addition, $q^0$, the energy carried by particle 3 in the rest frame of the three particle system, is given by 
\be 
q^0(s)=\frac{s+m_3^2-M_R^2}{2\sqrt{s}}.
\ee  

   Note also that the arguments of the amplitudes $T_i(s)$ and $t_i(s_i)$ are different, where $s$ is the total invariant mass of the three-body system, and $s_i$ are the invariant masses in the two-body systems. The value of $s_i$ is given by \cite{Yamagata2010}
\be 
s_i=m_3^2+m_i^2+\frac{(M_R^2+m_i^2-m_j^2)(s-m_3^2-M_R^2)}{2M_R^2}, (i,j=1,2,\;i\neq j)
\ee
where $m_l, (l=1,2,3)$ are the masses of the corresponding particles in the three-body system and $M_R$ the mass of two body resonance or bound state (cluster).

\section{Results}

The $D^*$-multi-$\rho$ interactions that we investigate in the present work are listed in Table \ref{cases}, and are explained as follows. For the three-body interaction, we have two options: particle $3=D^*$, cluster or resonance $R=f_2$ (particle $1=\rho,\;2=\rho$) and $3=\rho$, $R=D^*_2$ ($1=\rho,\;2=D^*$). For four-body, we also have two cases: $3=f_2$, $R=D^*_2$ ($1=\rho,\;2=D^*$) and $3=D^*_2$, $R=f_2$ ($1=\rho,\;2=\rho$). For five-body, $3=D^*$, $R=f_4$ ($1=f_2,\;2=f_2$) and $3=\rho$, $R=D^*_4$ ($1=f_2,\;2=D^*_2$). For six-body, $3=D^*_2$, $R=f_4$ ($1=f_2,\;2=f_2$) and $3=f_2$, $R=D^*_4$ ($1=f_2,\;2=D^*_2$). We describe all these cases in detail below.
\begin{table}[htb]
\centering
\caption{The cases considered in the $D^*$-multi-$\rho$ interactions.}\label{cases}
\begin{tabular}{cccc}
\hline\hline
\hspace{0.3cm} particles: \hspace{0.3cm} & \hspace{0.5cm} 3 \hspace{0.5cm} & \hspace{0.5cm} R (1,2) \hspace{0.5cm} &  \hspace{0.3cm} amplitudes \hspace{0.3cm} \\
\hline
Two-body & $\rho$ & $D^*$ &  $t_{\rho D^*}$ \\
  & $\rho$ & $\rho$ & $t_{\rho\rho}$  \\
\hline
Three-body & $D^*$ & $f_2\;(\rho\rho)$ & $T_{D^*-f_2}$  \\
  & $\rho$ & $D^*_2\;(\rho D^*)$ & $T_{\rho-D^*_2}$  \\
\hline
Four-body  & $D^*_2$ & $f_2\;(\rho\rho)$ & $T_{D^*_2-f_2}$  \\
  & $f_2$ & $D^*_2\;(\rho D^*)$ & $T_{f_2-D^*_2}$ \\
\hline
Five-body & $D^*$ & $f_4\;(f_2 f_2)$ & $T_{D^*-f_4}$  \\
  & $\rho$ & $D^*_4\;(f_2 D^*_2)$ & $T_{\rho-D^*_4}$  \\
\hline
Six-body  & $D^*_2$ & $f_4\;(f_2 f_2)$ & $T_{D^*_2-f_4}$  \\
  & $f_2$ & $D^*_4\;(f_2 D^*_2)$ & $T_{f_2-D^*_4}$ \\
\hline\hline
\end{tabular}
\end{table}

\subsection{Three-body interaction}
\label{threeb}

For three-body interaction, we have two options of structure: $D^*-f_2(\rho\rho)$ and $\rho-D^*_2(\rho D^*)$, which means $3=D^*$, $R=f_2$ ($1=\rho,\;2=\rho$) and $3=\rho$, $R=D^*_2$ ($1=\rho,\;2=D^*$). Thus, to evaluate these scattering amplitudes, we need as input the $t_1$ and $t_2$ amplitudes of the (3,1) and (3,2) systems, $t_1 = t_2 = t_{\rho D^*}$ for $D^*-f_2(\rho\rho)$ and $t_1 = t_{\rho\rho},\; t_2=t_{\rho D^*}$ for $\rho-D^*_2(\rho D^*)$. We should calculate the two-body $\rho\rho$ and $\rho D^*$ amplitudes.

As mentioned before, the isospin structure of the cluster should be considered for the $t_1$ and $t_2$ amplitudes. For the case of $D^*-f_2(\rho\rho)$, the cluster of $f_2$ has isospin $I=0$. Therefore the two $\rho$ mesons are in an $I=0$ state, and we have
\be 
|\rho\rho>^{(0,0)}=\frac{1}{\sqrt{3}} \Big( |(1,-1)> + |(-1,1)> - |(0,0)> \Big),
\ee
where $|(1,-1)>$ denote $|(I_z^1,I_z^2)>$ which shows the $I_z$ components of particles 1 and 2, and $|\rho\rho>^{(0,0)}$ means $|\rho\rho>^{(I,I_z)}$. The third particle is a $D^*$ meson taken $|I_z^3)>=|\frac{1}{2}>$. Then we obtain
\be 
\begin{split}
T^{(\frac{1}{2},\frac{1}{2})}_{D^*-f_2}=&<D^* \rho\rho|\,\hat{t}\,|D^* \rho\rho>^{(\frac{1}{2},\frac{1}{2})}  \\
=&(<D^*|^{(\frac{1}{2},\frac{1}{2})} \otimes <\rho\rho|^{(0,0)})\,(\hat{t}_{31}+\hat{t}_{32})\,(|D^*>^{(\frac{1}{2},\frac{1}{2})} \otimes |\rho\rho>^{(0,0)})\\
=&\Big[ <\frac{1}{2}| \otimes \frac{1}{\sqrt{3}} \Big( <(1,-1)| + <(-1,1)| - <(0,0)| \Big) \Big]\,(\hat{t}_{31}+\hat{t}_{32})\,\Big[ |\frac{1}{2}> \\
&\otimes \frac{1}{\sqrt{3}} \Big( |(1,-1)> + |(-1,1)> - |(0,0)> \Big) \Big]\\
=&\frac{1}{3} \Big[ <(\frac{3}{2},\frac{3}{2}),-1| + \sqrt{\frac{1}{3}} <(\frac{3}{2},-\frac{1}{2}),1| + \sqrt{\frac{2}{3}} <(\frac{1}{2},-\frac{1}{2}),1| - \sqrt{\frac{2}{3}} <(\frac{3}{2},\frac{1}{2}),0| \\
&- \sqrt{\frac{1}{3}} <(\frac{1}{2},\frac{1}{2}),0| \Big] \hat{t}_{31} \Big[ |(\frac{3}{2},\frac{3}{2}),-1> + \sqrt{\frac{1}{3}} |(\frac{3}{2},-\frac{1}{2}),1> + \sqrt{\frac{2}{3}} |(\frac{1}{2},-\frac{1}{2}),1> \\
&- \sqrt{\frac{2}{3}} |(\frac{3}{2},\frac{1}{2}),0> - \sqrt{\frac{1}{3}} |(\frac{1}{2},\frac{1}{2}),0> \Big] +\frac{1}{3} \Big[ \sqrt{\frac{1}{3}} <(\frac{3}{2},-\frac{1}{2}),1| + \sqrt{\frac{2}{3}} <(\frac{1}{2},-\frac{1}{2}),1| \\
&+ <(\frac{3}{2},\frac{3}{2}),-1| - \sqrt{\frac{2}{3}} <(\frac{3}{2},\frac{1}{2}),0| - \sqrt{\frac{1}{3}} <(\frac{1}{2},\frac{1}{2}),0| \Big] \hat{t}_{32} \Big[  \sqrt{\frac{1}{3}} |(\frac{3}{2},-\frac{1}{2}),1> \\
&+ \sqrt{\frac{2}{3}} |(\frac{1}{2},-\frac{1}{2}),1> + |(\frac{3}{2},\frac{3}{2}),-1> - \sqrt{\frac{2}{3}} |(\frac{3}{2},\frac{1}{2}),0> - \sqrt{\frac{1}{3}} |(\frac{1}{2},\frac{1}{2}),0> \Big],\label{tDsf2}
\end{split}
\ee
where the notation of the states followed in the terms is $|(\frac{3}{2},\frac{3}{2}),-1> \equiv |(I^{31},I_z^{31}),I_z^2>$ for $t_{31}$, and $|(I^{32},I_z^{32}),I_z^1>$ for $t_{32}$. Then we find
\be
t_1 = t_{\rho D^*} = \frac{1}{3} \big( 2 t_{31}^{I=3/2} + t_{31}^{I=1/2} \big),\quad t_2 = t_1.\label{tDsrho}
\ee

But for the case of $\rho-D^*_2(\rho D^*)$, the situation is different. Because the isospins of $\rho$ and $D^*_2$ are $I_\rho=1$ and $I_{D^*_2}=\frac{1}{2}$, the total isospin of the three-body system are $I_{total}\equiv I_{\rho\rho D^*}=\frac{1}{2}$ or $I_{total}\equiv I_{\rho\rho D^*}=\frac{3}{2}$, and then we have
\be 
\begin{split}
&|\rho D^*_2>^{(\frac{1}{2},\frac{1}{2})} = |\rho\rho D^*>^{(\frac{1}{2},\frac{1}{2})} = \sqrt{\frac{2}{3}} |(1,-\frac{1}{2})> - \sqrt{\frac{1}{3}} |(0,\frac{1}{2})>, \\
&|\rho D^*_2>^{(\frac{3}{2},\frac{1}{2})} = |\rho\rho D^*>^{(\frac{3}{2},\frac{1}{2})} = \sqrt{\frac{1}{3}} |(1,-\frac{1}{2})> + \sqrt{\frac{2}{3}} |(0,\frac{1}{2})>,\label{trhoDs1}
\end{split}
\ee
where we have taken $I_z = \frac{1}{2}$ for convenience. Therefore the $|\rho D^*>$ states inside the $D^*_2$ for the $I_z = -\frac{1}{2}$ and $I_z = +\frac{1}{2}$ are given by
\be 
\begin{split}
&|\rho D^*>^{(\frac{1}{2},-\frac{1}{2})} = \sqrt{\frac{1}{3}} |(0,-\frac{1}{2})> - \sqrt{\frac{2}{3}} |(-1,\frac{1}{2})>, \\
&|\rho D^*>^{(\frac{1}{2},\frac{1}{2})} = \sqrt{\frac{2}{3}} |(1,-\frac{1}{2})> - \sqrt{\frac{1}{3}} |(0,\frac{1}{2})>.\label{trhoDs2}
\end{split}
\ee
For the two possibilities, combining Eqs. \eqref{trhoDs1} and \eqref{trhoDs2} and performing a similar derivation of Eq. \eqref{tDsf2}, we obtain
\be 
\begin{split}
&T_{\rho-D^*_2}^{(I=1/2)}: \quad t_1 = t_{\rho\rho} = \frac{2}{3} t_{31}^{(I=0)}, \quad t_2 = t_{\rho D^*} = \frac{1}{9} \big( 8 t_{32}^{I=3/2} + t_{32}^{I=1/2} \big); \\
&T_{\rho-D^*_2}^{(I=3/2)}: \quad t_1 = t_{\rho\rho} = \frac{5}{6} t_{31}^{(I=2)}, \quad t_2 = t_{\rho D^*} = \frac{1}{9} \big( 5 t_{32}^{I=3/2} + 4 t_{32}^{I=1/2} \big).
\end{split}
\ee

\begin{figure}
\centering
\includegraphics[scale=0.6]{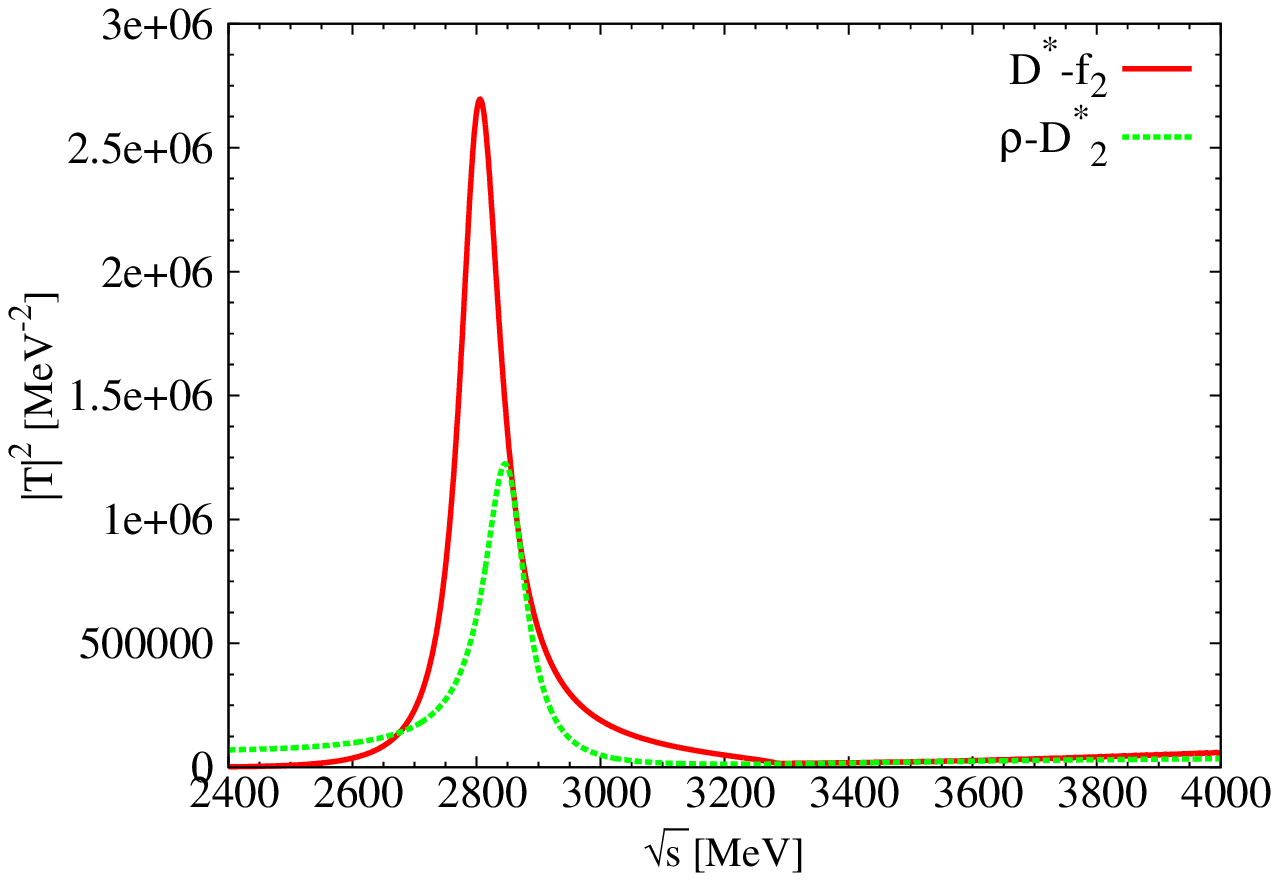}
\includegraphics[scale=0.6]{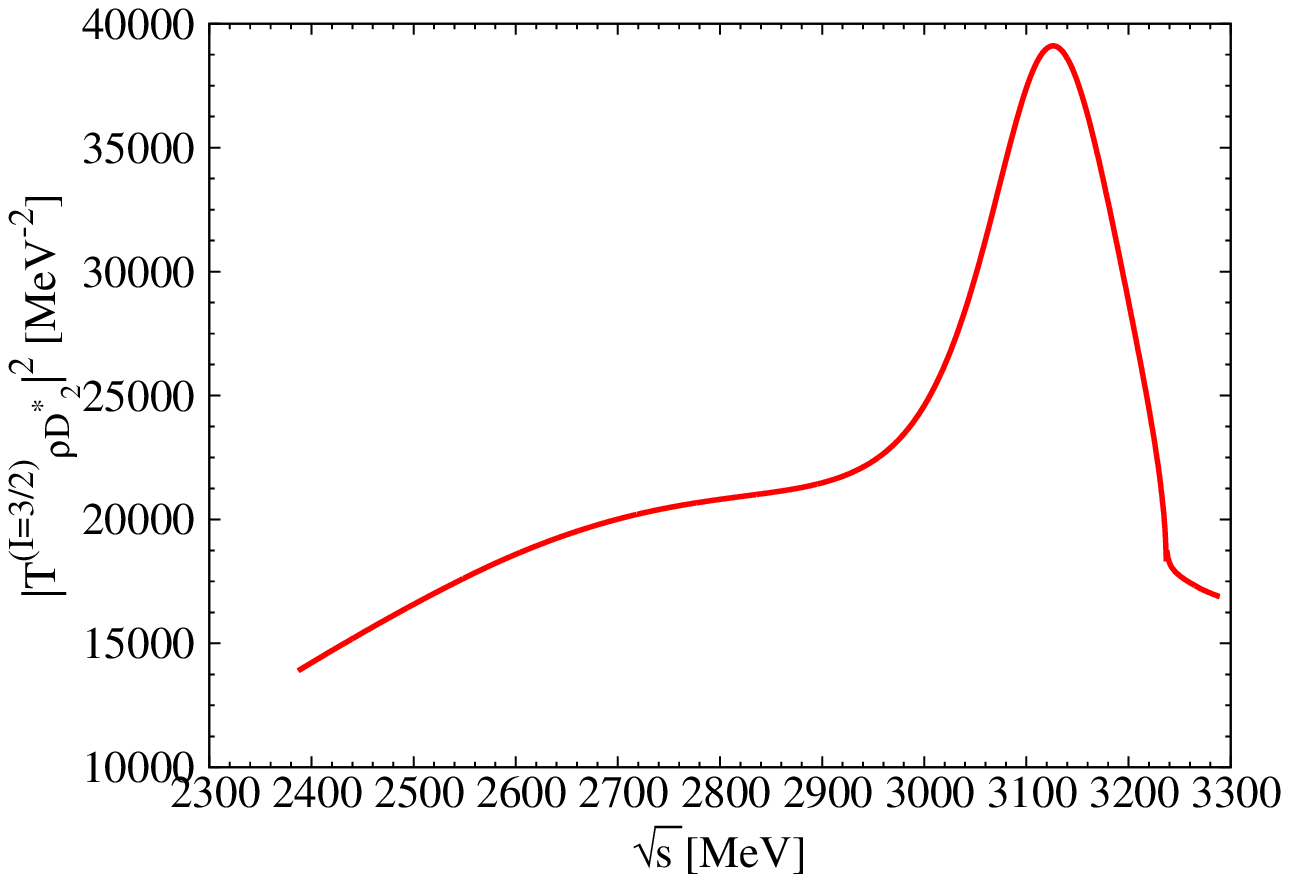}
\caption{Modulus squared of the $T_{D^*-f2}$ and $T_{\rho-D^*_2}$ scattering amplitudes. Left: $I_{total}=\frac{1}{2}$; Right: $I_{total}=\frac{3}{2}$.} \label{fig1}
\end{figure}

We show our results in Fig. \ref{fig1}. In Fig. \ref{fig1} (left) we show the modulus squared of the amplitudes for $|T_{D^*-f_2}^{I=1/2}|^2$ and $|T_{\rho-D^*_2}^{I=1/2}|^2$, and we find that there are clear peaks around the energy $2800-2850\mev$ which is about $400\mev$ lower than the $D^*-f_2$ threshold. The bindings are large because they scale with the mass of the mesons and we have now a $D^*$ interacting with two $\rho$ mesons. The strength of the peak of $|T_{D^*-f_2}^{I=1/2}|^2$ is two times bigger than for $|T_{\rho-D^*_2}^{I=1/2}|^2$, and we see that the $D^*-f_2$ component is a bit more bound than the $\rho-D^*_2$ one. We expect that a real state would be an admixture of both with a binding in between that of the individual components. In Fig. \ref{fig1} (right) we show $|T_{\rho-D^*_2}^{I=3/2}|^2$, and there is a clear resonant structure about $3120\mev$, the strength of which is 30 times smaller than that of $|T_{\rho-D^*_2}^{I=1/2}|^2$ in the left figure and less bound. We are concerned with the lowest lying states and hence we concentrate on the predicted new $D^*_3$ state with a structure formed by a mixture of $D^*-f_2$ and $\rho-D^*_2$, with a mass about $2800-2850\mev$ and a width about $60-100\mev$.

\subsection{Four-body interaction}
\label{fourb}

There are also two possibilities in the four-body interaction as we have shown in Table \ref{cases}: particle $3=f_2$, $R=D^*_2$ ($1=\rho,\;2=D^*$) or particle $3=D^*_2$, $R=f_2$ ($1=\rho,\;2=\rho$). Because $I_{f_2} = 0$ and $I_{D^*_2} = \frac{1}{2}$, the total isospin of the four-body system is only $I_{total} = \frac{1}{2}$. In the first case, $f_2$ collides with the $D^*_2$, the amplitudes $t_1 = t_{f_2 \rho} = T_{\rho-f_2}$ has been evaluated in Ref. \cite{Roca2010} and is reproduced in our work, and $t_2 = t_{f_2 D^*} = T_{D^*-f_2}$ which has been evaluated in the former subsection \ref{threeb}. For the second case, $D^*_2$ collides with the $f_2$, and the amplitudes $t_1 = t_2 = t_{D^*_2 \rho} = T_{\rho-D^*_2}$ have been evaluated in the former subsection \ref{threeb}. We must now consider that the three-body amplitude $T_{\rho-D^*_2}$ is also combined with different isospins as mentioned in subsection \ref{threeb}. This situation is similar to the case when the $D^*$ collides with the $f_2$, because the isospins of both the $D^*_2$ and $D^*$ are $I=\frac{1}{2}$, thus from Eq. \eqref{tDsrho} we have
\be
t_1 = T_{\rho D^*_2} = \frac{1}{3} \big( 2 T_{31}^{I=3/2} + T_{31}^{I=1/2} \big),\quad t_2 = t_1.\label{threet3}
\ee

\begin{figure}
\centering
\includegraphics[scale=0.6]{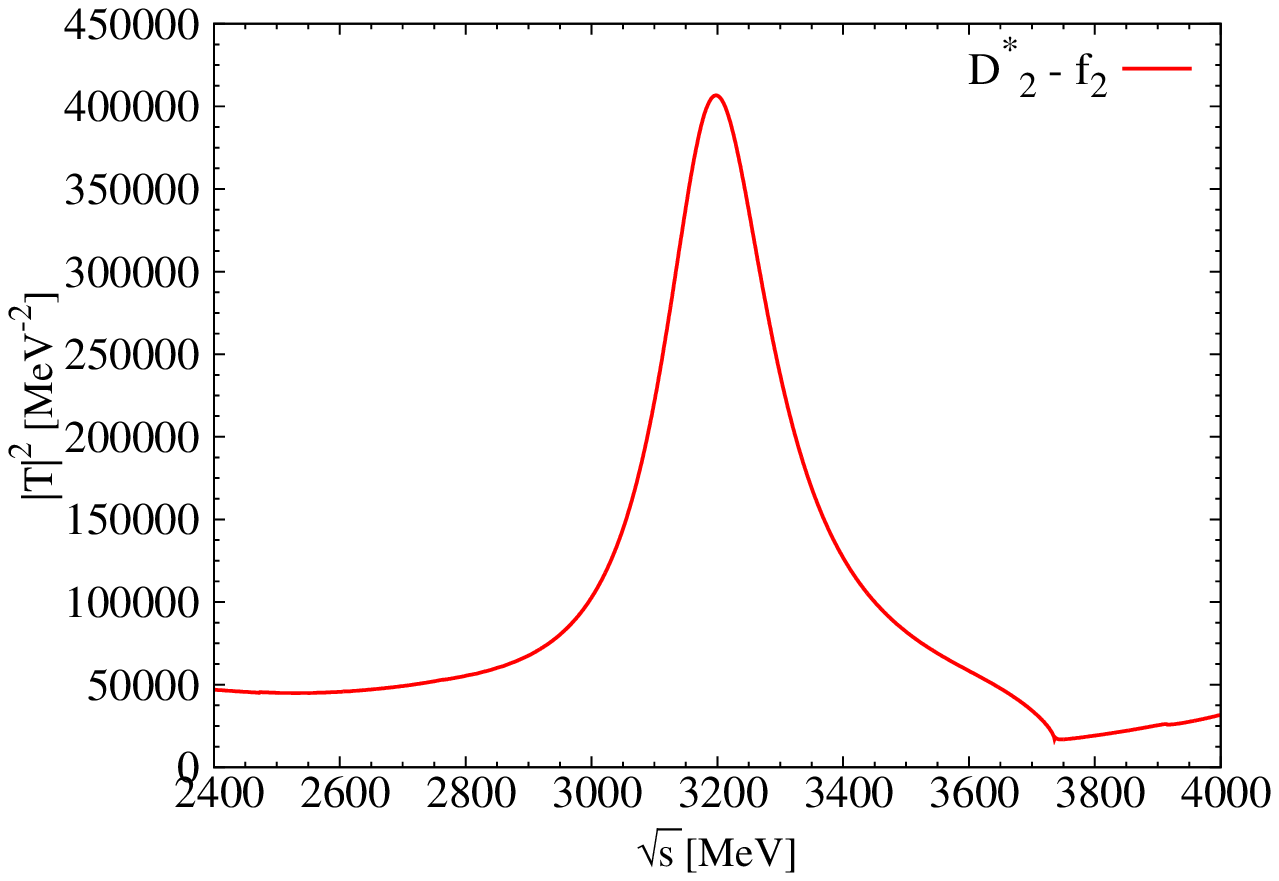}
\includegraphics[scale=0.6]{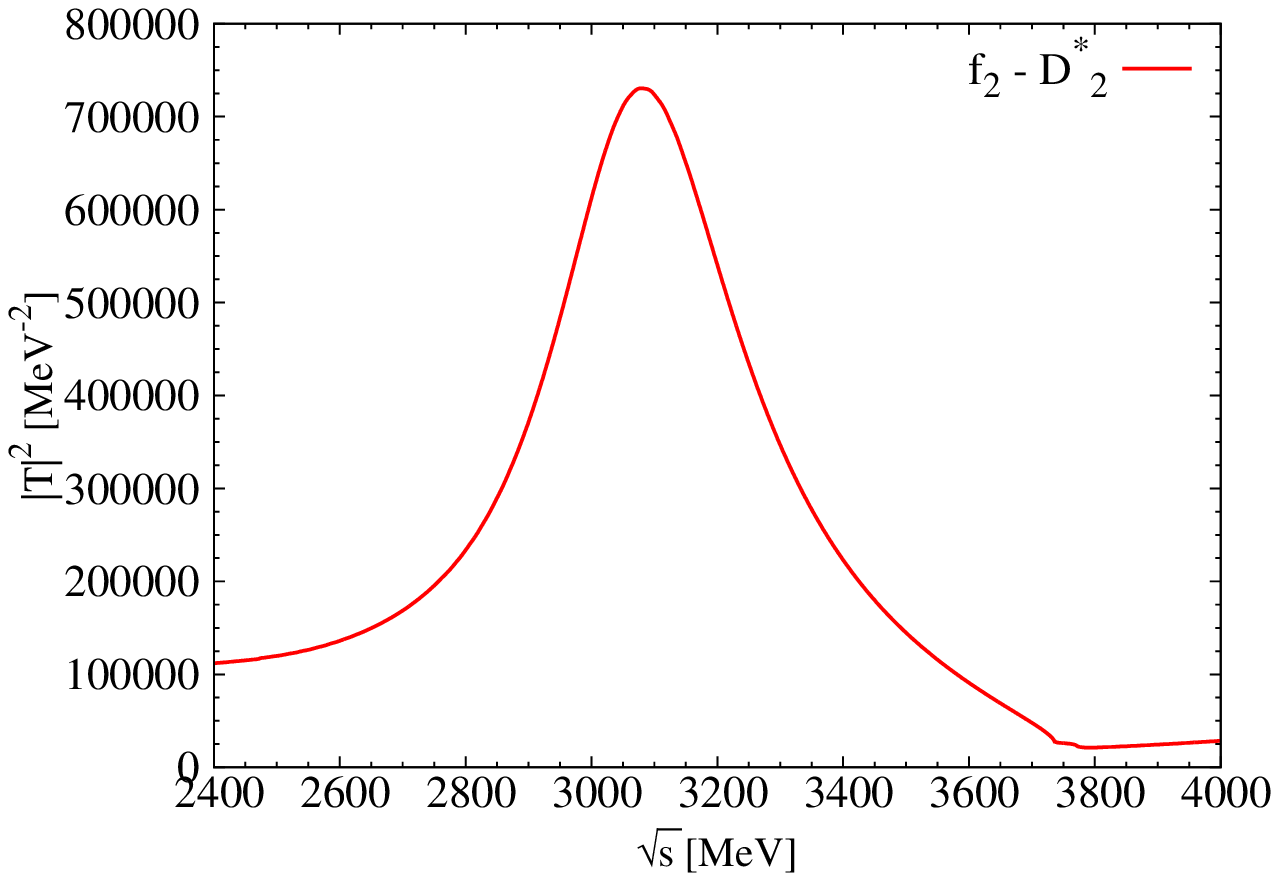}
\caption{Modulus squared of the $T_{D^*_2-f2}$ (left) and $T_{f_2-D^*_2}$ (right) scattering amplitudes.} \label{fig2}
\end{figure}

The results are shown in Fig. \ref{fig2}. The left of Fig. \ref{fig2} is $|T_{D^*_2-f_2}^{I=1/2}|^2$. We find that there is a clear peak at an energy of $3200\mev$, the width of which is about $200\mev$. The right of Fig. \ref{fig2} shows $|T_{f_2-D^*_2}^{I=1/2}|^2$ and there is a resonant peak around the energy $3075\mev$ with a large width of nearly $400\mev$. The strength of the peak of $|T_{f_2-D^*_2}^{I=1/2}|^2$ is about two times bigger than the one of $|T_{D^*_2-f_2}^{I=1/2}|^2$ and the energy of the peak is more bound too. But from the former results, subsection \ref{threeb}, we found that $|T_{D^*_2-f_2}^{I=1/2}|^2$ has more strength and is more bound than $|T_{f_2-D^*_2}^{I=1/2}|^2$. We have investigated that this is because of the contribution of $|T_{\rho-D^*_2}^{I=3/2}|^2$, even though the strength of $|T_{\rho-D^*_2}^{I=3/2}|^2$ is much smaller and less bound than the one of $|T_{\rho-D^*_2}^{I=1/2}|^2$ from the former results. When we removed the contribution of $|T_{\rho-D^*_2}^{I=3/2}|^2$ in Eq. \eqref{threet3}, the strength of the peak was enhanced by a factor five and was more bound. Therefore, within the uncertainty of the theory, we find a new $D^*_4$ resonance, of a mass about $3075-3200\mev$ and a width about $200-400\mev$.

\subsection{Five-body interaction}

For the five-body interaction, we also have two options for the cluster, one of which is the particle $f_4$ studied in Ref. \cite{Roca2010} and the other one the resonance $D^*_4$ obtained in the four-body interaction, subsection \ref{fourb}. Thus letting the third particle ($D^*$ or $\rho$) collide with them, we have $3=D^*$, $R=f_4$ ($1=f_2,\;2=f_2$) or $3=\rho$, $R=D^*_4$ ($1=f_2,\;2=D^*_2$). Because the isospin $I_{f_4} = 0$ and $I_{D^*_4} = \frac{1}{2}$, the total isospin of the five-body system is only $I_{total} = \frac{1}{2}$ in the $D^*-f_4$ structure, but $I_{total} = \frac{1}{2}$ or $I_{total} = \frac{3}{2}$ in the $\rho-D^*_4$ structure. Thus the situation is similar to the three-body interaction discussed before, $D^*$ (or $\rho$) collide with $f_2$ (or $D^*_2$). Therefore in the first case, the $D^*$ collides with the $f_4$, and the amplitudes $t_1 = t_2 = t_{D^* f_2} = T_{D^*-f_2}^{(I=1/2)}$ have been evaluated in subsection \ref{threeb}. For the second case, the $\rho$ collides with the $D^*_4$, which is similar to $\rho-D^*_2$ in the three-body interaction, thus, after doing a similar derivation as in Eq. \eqref{tDsf2}, we have
\be 
\begin{split}
&T_{\rho-D^*_4}^{(I=1/2)}: \quad t_1 = t_{\rho f_2} = T_{31}^{(I=1)}, \quad t_2 = t_{\rho D^*_2} =  T_{32}^{I=1/2}; \\ 
&T_{\rho-D^*_4}^{(I=3/2)}: \quad t_1 = t_{\rho f_2} = T_{31}^{(I=1)}, \quad t_2 = t_{\rho D^*_2} =  T_{32}^{I=3/2},
\end{split}
\ee
where the $T_{31}^{(I=1)}$ is the same as $T_{\rho-f_2}$ in the subsection \ref{fourb} reproducing the results of Ref. \cite{Roca2010}, and $T_{\rho-D^*_2}^{I=1/2}$ and $T_{\rho-D^*_2}^{I=3/2}$ have also been evaluated in subsection \ref{threeb}.

\begin{figure}
\centering
\includegraphics[scale=0.6]{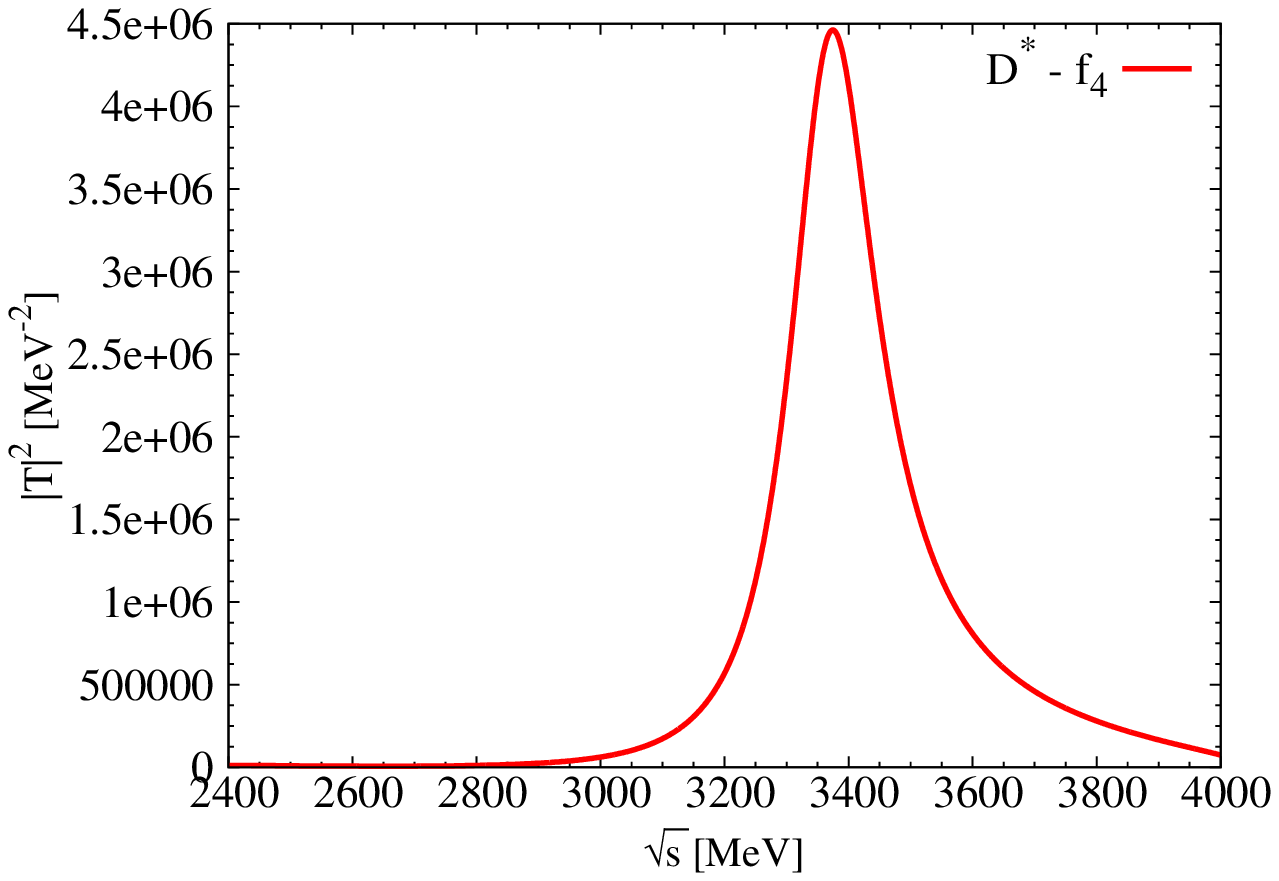}
\includegraphics[scale=0.6]{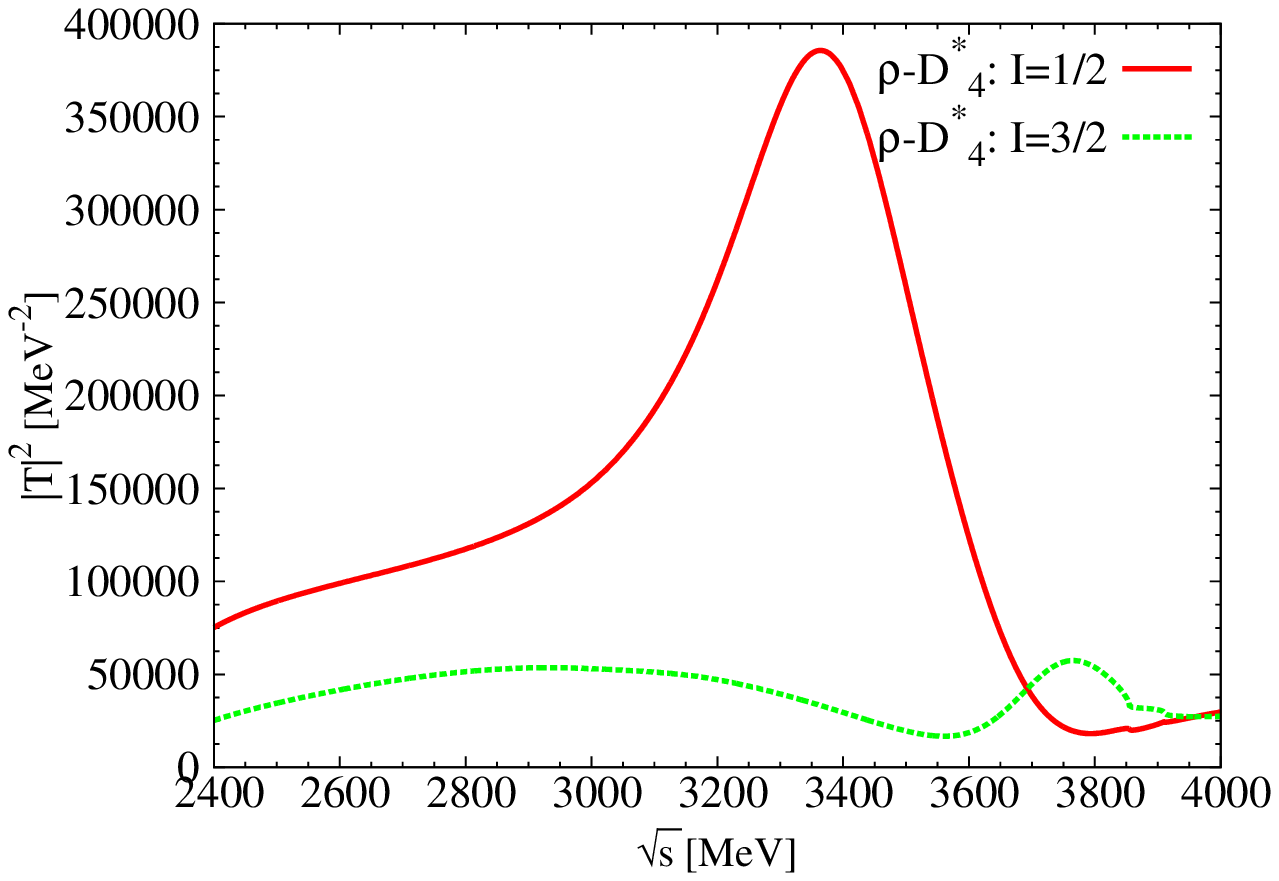}
\caption{Modulus squared of the $T_{D^*-f_4}$ (left) and $T_{\rho-D^*_4}$ (right) scattering amplitudes.} \label{fig3}
\end{figure}

In Fig. \ref{fig3} we show our results. The left of Fig. \ref{fig3} is $|T_{D^*-f_4}^{I=1/2}|^2$ and we observe a resonant peak around the energy $3375\mev$ with a width of less than $200\mev$. The right of Fig. \ref{fig3} is $|T_{\rho-D^*_4}^{I=1/2}|^2$ and $|T_{\rho-D^*_4}^{I=3/2}|^2$. We find that there is a resonant structure in $|T_{\rho-D^*_4}^{I=1/2}|^2$ at the energy $3360\mev$, the width of which is about $400\mev$, and the position is very close to the one of $|T_{D^*-f_4}^{I=1/2}|^2$. But the strength of $|T_{\rho-D^*_4}^{I=1/2}|^2$ is one order smaller than the one $|T_{D^*-f_4}^{I=1/2}|^2$. For $|T_{\rho-D^*_4}^{I=3/2}|^2$ there is no resonant structure. Therefore, within uncertainties, we also find a new $D^*_5$ resonance, with a mass about $3360-3375\mev$ and a width about $200-400\mev$.

\subsection{Six-body interaction}

Similarly to the five-body interaction, we also have two options of the cluster for the six-body interaction, the particle $f_4$ studied in Ref. \cite{Roca2010} and the resonance $D^*_4$ obtained in subsection \ref{fourb}. Now letting a resonance ($D^*_2$ or $f_2$) be the third particle and collide with them, we have $3=D^*_2$, $R=f_4$ ($1=f_2,\;2=f_2$) or $3=f_2$, $R=D^*_4$ ($1=f_2,\;2=D^*_2$). Because $I_{f_2} = I_{f_4} = 0$ and $I_{D^*_2} = I_{D^*_4} = \frac{1}{2}$, the total isospin of the six-body system is only $I_{total} = \frac{1}{2}$. Thus, in the first case, the $D^*_2$ collides with the $f_4$, the amplitudes $t_1 = t_2 = t_{D^*_2 f_2} = T_{D^*_2-f_2}^{(I=1/2)}$ have been evaluated in subsection \ref{fourb}. For the second case, the $f_2$ collides with the $D^*_4$, the amplitudes $t_1 = t_{f_2 f_2} = T_{f_2-f_2}$ reproduce the results from Ref. \cite{Roca2010}, and $t_2 = t_{f_2 D^*_2} = T_{f_2-D^*_2}$ has been calculated in subsection \ref{fourb}. 

\begin{figure}
\centering
\includegraphics[scale=0.6]{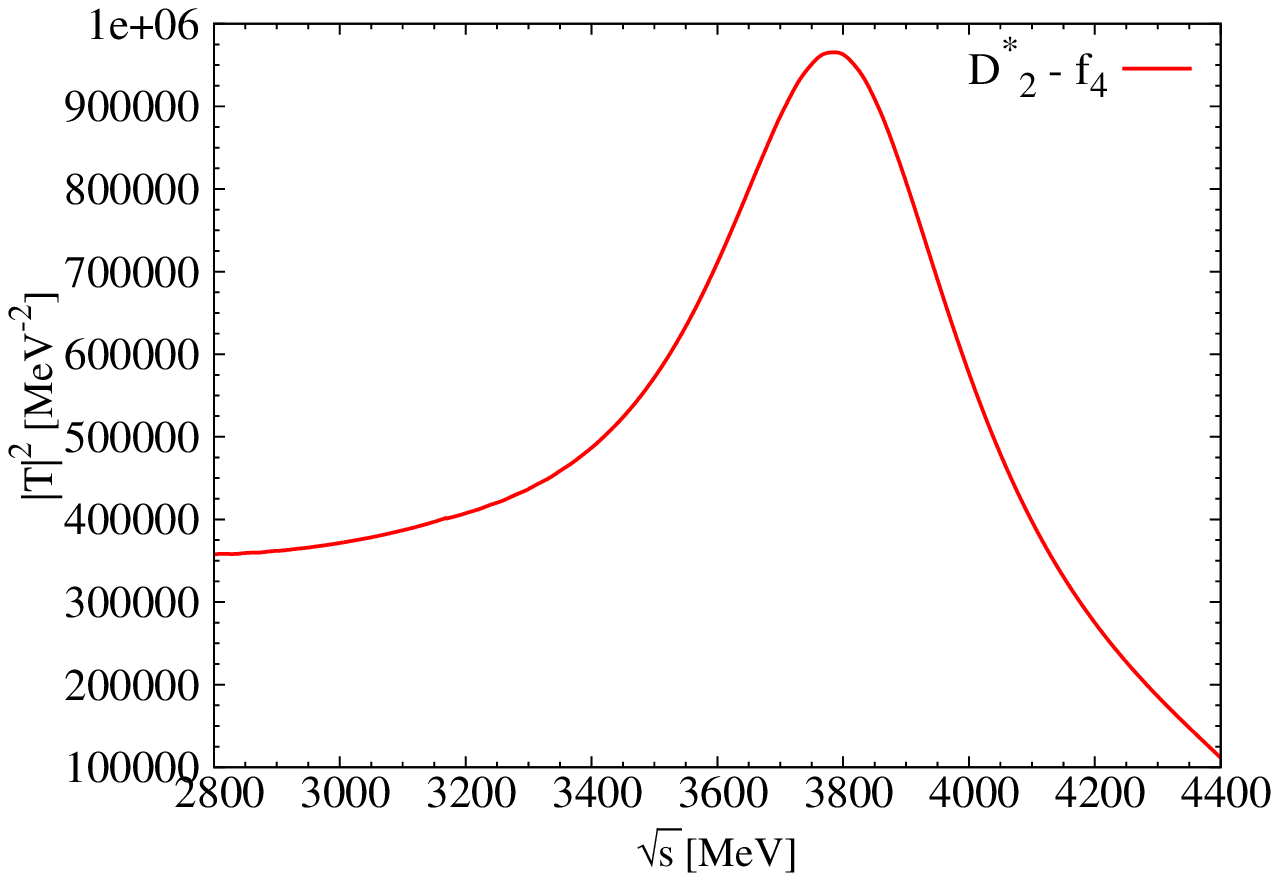}
\includegraphics[scale=0.6]{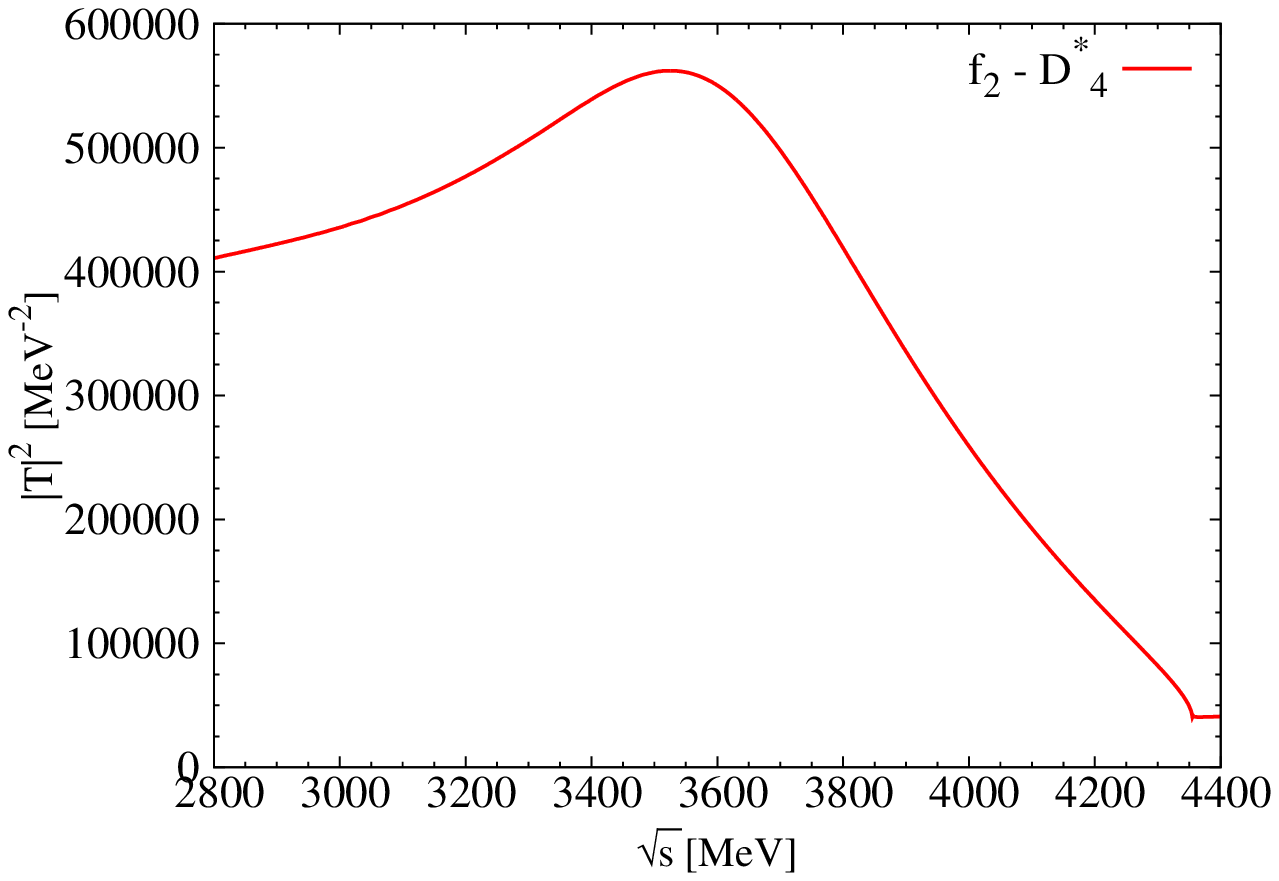}
\caption{Modulus squared of the $T_{D^*_2-f_4}$ (left) and $T_{f_2-D^*_4}$ (right) scattering amplitudes.} \label{fig4}
\end{figure}

Our results are shown in Fig. \ref{fig4}. The left of Fig. \ref{fig4} is $|T_{D^*_2-f_4}^{I=1/2}|^2$ where we see a peak around the energy $3775\mev$ with a large width of nearly $400\mev$. The right of Fig. \ref{fig4} is $|T_{f_2-D^*_4}^{I=1/2}|^2$, and there we find that there is not a clear peak at the energy $3550\mev$. It looks like the resonant structure of $f_2-D^*_4$ is not as stable as the $D^*_2-f_4$ one. From these results, we could predict a new $D^*_6$ resonance with more uncertainty, with a mass of about $3775\mev$ and a width about $400\mev$.

\section{Conclusions}

In the present work, we show the results of our investigation of the $D^*$-multi-$\rho$ systems. Our idea is based on the fact that the two-body interactions of $\rho\rho$ and $\rho D^*$ in spin $S=2$ are so strong as to bind the particles forming the resonances$f_2(1270)$ \cite{Molina:2008jw} and $D^*_2(2460)$ \cite{Molina:2009eb} respectively. So we could study the many-body $D^*$-multi-$\rho$ systems in an iterative way looking at the structure of the amplitudes and observing clear peaks that become wider as the number of $\rho$ mesons increase. The work proceeded analogously to the study of multi-$\rho$ system in \cite{Roca2010}, where the $\rho_3(1690)(3^{--})$, $f_4(2050)(4^{++})$, $\rho_5(2350)(5^{--})$, and $f_6(2510)(6^{++})$ were described as basically molecules of multi-$\rho(770)$ states, and similarly to the work of \cite{Yamagata2010} where the $K^*_2(1430)$, $K^*_3(1780)$, $K^*_4(2045)$, $K^*_5(2380)$ and $K^*_6$ could be interpreted as molecules made of one $K^*(892)$ meson and an increasing number of $\rho(770)$ mesons. The $D^*$-multi-$\rho$ states with spins aligned combined to give some new charmed resonances, $D^*_3$, $D^*_4$, $D^*_5$ and $D^*_6$, which are basically made of one $D^*$ meson and an increasing number of $\rho(770)$ mesons and are not found in the list of PDG \cite{pdg2012}. Their masses are predicted around $2800-2850\mev$, $3075-3200\mev$, $3360-3375\mev$ and $3775\mev$ respectively. And their widths are about $60-100\mev$, $200-400\mev$, $200-400\mev$ and $400\mev$ respectively. The analogy with the states already known in the strange and non-strange sector, together with the stronger interaction of the $D^*$ mesons, make our predictions solid within the uncertainties admitted. We are, thus, reasonably confident that such states can be found in the future in coming facilities like FAIR and others.

\section*{Acknowledgements}

We thank R. Molina for useful discussions.
This work is partly supported by DGICYT contract number FIS2011-28853-C02-01, and the Generalitat Valenciana in the program Prometeo, 2009/090. We acknowledge the support of the European Community-Research Infrastructure Integrating Activity Study of Strongly Interacting Matter (acronym HadronPhysics3, Grant Agreement n. 283286) under the Seventh Framework Programme of the EU.
One of us, M. Bayar acknowledges support through the Scientific and Technical Research Council (TUBITAK) BIDEP-2219 grant.

\end{document}